\documentclass{PoS}

\title{Reactor Antineutrinos Signal all over the world}

\ShortTitle{Reactor antineutrinos signal all over the world}

\author{\speaker{B. Ricci}, F.Mantovani, M. Baldoncini 
\\
          Dipartimento di Fisica e Scienze della Terra, 
          Universit\`a and INFN,  Ferrara, I-44122 Italy\\
        E-mail: \email{ricci@fe.infn.it}}

\author{J. Esposito\\
        INFN, Laboratori Nazionali di Legnaro, Padova I-35020- Italy\\
        E-mail: \email{esposito@lnl.infn.it}}

\author{L. Ludhova\\
          Dipartimento di Fisica, Universit\`a and INFN, Milano I-20133, Italy \\
        E-mail: \email{ludhova@gmail.com}}

\author{S. Zavatarelli\\
          Dipartimento di Fisica, Universit\`a and INFN, Genova I-16146, Italy \\
        E-mail: \email{zavatarelli@ge.infn.it}}

\abstract{
We present an updated estimate of reactor antineutrino 
signal all over the world, with  particular attention to the sites 
proposed for existing and future geo-neutrino experiment. 
In our calculation we take into account the most updated data
on Thermal Power for each nuclear plant, on reactor antineutrino spectra
and on three neutrino oscillation mechanism.}

\FullConference{XV Workshop on Neutrino Telescopes,\\
		11-15 March 2013\\
		Venice, Italy}

\begin{document}

\begin{table}
\center
\begin{tabular}{|l|c|c|c|c|}
\hline 
 Sites & R [TNU] &  R$_G$ [TNU] & G [TNU] & R$_G$/G \\
\hline
  LNGS &  85.8	       $\pm$	4.6   &  22.8	$\pm$	1.1   &   $40.3^{+7.3}_{-5.8}$  & 0.6 \\ 
\hline
  KAMIOKA & 70.1	$\pm$	3.7  & 18.7	$\pm$	1.1  &    $31.5^{+4.9}_{-4.1}$  & 0.6\\ 
\hline
  SUDBURY &  174.6	$\pm$	9.0  &  43.1	$\pm$	2.1   &  $45.4^{+7.5}_{-6.3}$  & 0.9\\ 
\hline
  PHYASALMI & 69.2	$\pm$	3.7 &  17.5	$\pm$	0.8 &    $45.3^{+7.0}_{-5.9}$  & 0.4\\ 
\hline
  FREJUS &  587.9	$\pm$	31.0  &  134.0	$\pm$	7.1 &    $42.4^{+7.6}_{-6.2}$  & 3.2\\ 
\hline
  HOMESTAKE & 27.7	$\pm$	1.5  & 7.3	$\pm$	0.3  &  $48.7^{+8.4}_{-6.9}$  & 0.1\\  
\hline
  HAWAII &  3.4	$\pm$	0.2  &  0.9	$\pm$	0.04 &           $12.0^{+0.7}_{-0.6}$  & 0.1\\ 
\hline
  CURACAO &  9.5	$\pm$	0.5 &  2.5	$\pm$	0.1  &  $29.3^{+4.2}_{-3.3}$  & 0.1\\
\hline
\end{tabular}
\caption{Comparison between expected reactor (R) and geo (G) antineutrino signal.
R$_G$ indicates the reactor signale expected in the geo neutrino energy window ($E_{\bar{\nu}}<3.26 $ MeV). 
Geo-neutrino signal  has been  calculated following the approach described in \cite{hao}. 
1 TNU = 1event/year/$10^{32}$ protons.}
\label{table}
\end{table}

\section{Discussion}
Antineutrinos from the decay chains  of $^{238}$U and $^{232}$Th existing 
in the Earth interior  (the so called geo-neutrinos)
 have been recently detected both by Kamland \cite{kamland} and by Borexino \cite{borexino} experiments.
 Future experiments for geo-neutrinos detection have been proposed (or starting) in  several
 location in the world (e.g. SNO+ in Canada \cite{SNO+}, Lena project in Europe \cite{lena} and 
Hawaii Anti-Neutrino Observatory\cite{hawaii}). 

The main source of background of such experiments is given by antineutrino produced
 by nuclear plants. These particles account for a signal almost always larger than geo-neutrinos one, 
see Table \ref{table}. So a detailed calculation of reactor  antineutrino 
flux in mandatory for an accurate measurements of geo-neutrinos. 

With this aim, we performed a calculation of reactor antineutrinos flux all over the world.
Previus analysis has been presented, for instance, in ref. \cite{report} and \cite{geoscience2010}.
Now we will show an updated estimate of reactor antineutrino signal, 
with  particular attention to the sites proposed for the new geo-neutrino experiments. 
In our calculation we take into account the most updated data on Thermal Power for each nuclear plant, 
on reactor antineutrino spectra and on three neutrino oscillation mechanism.
The expected reactor antineutrino signal has been calculated as follows:
\begin{equation}
N_{ev}= \epsilon  N_p  \tau
\sum_{r=1}^{N_{react}} 
\frac {P_{r}}{4 \pi L_{r}^{2}}  <LF_r>  
\int dE_{\bar{\nu}_e} 
\sum_{i=1}^4 
\frac {f_{i}}{E_{i}} \phi_{i}(E_{\bar{\nu}_e}) 
\sigma(E_{\bar{\nu}_e})
P_{ee}(E_{\bar{\nu}_e};\hat\theta, L_r)
\label{Eq:ReactorFlux}
\end{equation}
where $\epsilon$ is detector efficiency,
$N_p$ is the number of target proton, 
$\tau$ is period  of data taking,
index $r$ cycles over the $N$ reactors considered, 
$L_{r}$,  $P_{r}$ and  $<LF_r>$ 
are the distance,  the nominal thermal power  and the averaged Load Factor of reactor
$r$, respectively.
The index $i$ stands for the i-th spectral component in the set 
($^{235}$U, $^{238}$U, $^{239}$Pu, and $^{241}$Pu), $f_{i}$ is the power fraction of the component $i$, 
as reported in\cite{borexino},
$E_i$ is the average antineutrino energy per fission of the component $i$ \cite{huber}, 
$\phi(E_{\bar{\nu}})$ is the anti-neutrino flux per fission of the $i^{\rm th}$ component, 
as recently calculated in ref.\cite{mueller}, 
$\sigma(E_{\bar{\nu}})$ is the inverse beta decay cross section\cite{vissani}
and $P_{ee}$ is the survival probability of the reactor antineutrinos of energy 
$E_{\bar{\nu}}$ traveling the baseline $L_r$,depending on the mixing parameters $\hat\theta$.

In Eq. (\ref{Eq:ReactorFlux}) we assume a 100\% detection efficiency, for a detector containing 
 $10^{32}$ target protons and operating continuously for 1 year.
In particular we consider the  nuclear cores all over the world,
operating in the year 2012. Information on the nominal thermal power and monthly load factor for each 
nuclear cores originates from International Agency of Atomic Energy (IAEA)~\cite{iaea}.
Concerning survival probability, we assumed a three flavour vacuum oscillation mechanism  with $P_{ee}$ as in
 in ref.\cite{fiorentini2012}, and  mixing parameters from  ref.\cite{fogli2012}.

\begin{figure}
\center
\includegraphics[width=.9\textwidth]{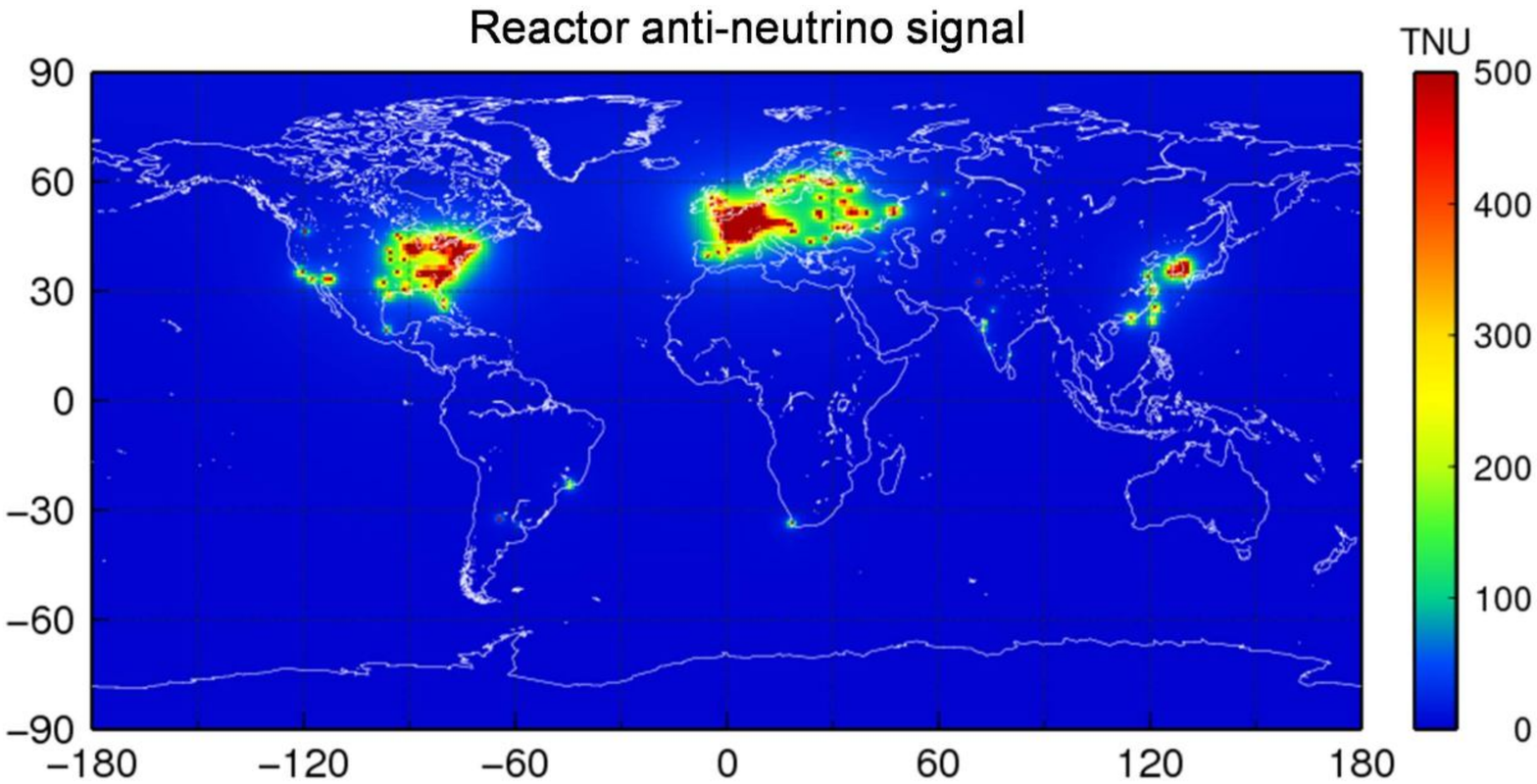}
\caption{A worldwide map of reactor antineutrinos signal. 1 TNU= 1 events/yr/10$^{32}$ target protons}
\label{mappa}
\end{figure}

The results of our calculation are reported in Table\ref{table} and in Fig.\ref{mappa}.
We also performed a  analysis on the sources of uncertainty in reactor signal prediction, see\cite{geoscience2010} for details. 
The total uncertainty is of the order of 5\%, the main contributions (i.e. greater than 2\%) 
arising from  $\theta_{12}$ mixing angle, antineutrino spectrum, fuel composition and thermal power.

One can see that, due to reactors shutdown occurred in 2012, 
Kamioka became a suitable site for detecting geo-neutrinos, comparable to LNGS. 
A new European geo-neutrino detector located at Frejus Laboratory  
requires a detailed  knowledge of closeby reactors;
 the choice of Phyasalmi looks better  in this respect.
Of course Hawaii and  Curacao are wonderful places for geo-neutrino studies due to their position 
far away from any nuclear plants of the world. The same holds for Homestake. 
In the near future, the SNO+ experiment, with a quite reasonable ratio 
R$_G$/G, will  provide  more information about Earth's interior.

\end{document}